**Deviation from secular equilibrium under laser exposure of gold nanoparticles in aqueous solutions of Uranium salt**

**A.V. Simakin and G.A. Shafeev**

Wave Research Center of A.M. Prokhorov General Physics Institute of the Russian Academy of Sciences, 38, Vavilov street, 119991 Moscow Russian Federation

Abstract

Laser exposure of gold nanoparticles in aqueous solutions of Uranium salt leads to accelerated decay of U238 nuclei and significant deviation from secular equilibrium. The samples demonstrate the enhanced gamma emission in the range of 54 keV during laser exposure.

The enhanced gamma-activity of solutions of $Th(NO_3)_4$ with Au nanoparticles was detected under their exposure to laser radiation [1,2]. This activity was assigned to the emission of short-living lead isotope $^{212}Pb$. However, the energy resolution of the spectrometer was not sufficient to resolve separate peaks of gamma-photons.

The samples were the solutions of $UO_2Cl_2$ of natural isotope composition either in $H_2O$ or $D_2O$. The typical concentration of uranium salt was 50 mg/ml. 1 ml of this solution was mixed with 1 ml of the colloidal solution of Au nanoparticles (NPs) obtained by laser ablation of a bulk gold target in either in $H_2O$ or $D_2O$. Average size of Au NPs was about 20 nm at concentration of $10^{14}$ $cm^{-3}$. Two laser sources were used, either a visible range Cu vapor laser (wavelength of 510.6 and 578.2 nm and pulse duration of 10 ns) and the third harmonics of a Nd:YAG laser (wavelength of 355 nm and pulse duration of 150 ps). The estimated peak power in the solution was $6\times10^{11}$ and $10^{12}$ $W/cm^2$, respectively.

Gamma-emission from samples under their irradiation with a Cu vapor laser was recorded in real time using a scintillation gamma-spectrometer that was fixed near the working cell. Since the initial level of gamma-emission is comparable with the natural background, both the spectrometer and the working cell were placed into a 5 cm thick lead housing. The laser beam was directed to the working cell through a small aperture in the housing. The energy resolution of the scintillation spectrometer was 3 keV per channel.

A stationary semiconductor gamma-spectrometer with high resolution (Ortec-65195-P) was used for the absolute measurements of activity of nuclides in Becquerel/ml in the energy range from 0.06 to 1.5 MeV. The spectrometer was calibrated for most common nuclides, such as $^{234}Th$, $^{235}U$, etc. of Uranium branching. The activities of the nuclides and therefore their content in the samples were compared before and after laser exposure.

The low-energy part of the gamma-spectrum of the sample of Uranium salt with Au NPs in $D_2O$ is presented in Fig. 1. Sharp maximum at photon energy of 54 keV is observed during laser exposure.

This signal rapidly decays after the end of the laser exposure. To avoid possible electromagnetic influence of the laser power supply, the laser was still on while the beam itself was shut down. No signal at this photon energy is observed 16 hours later.

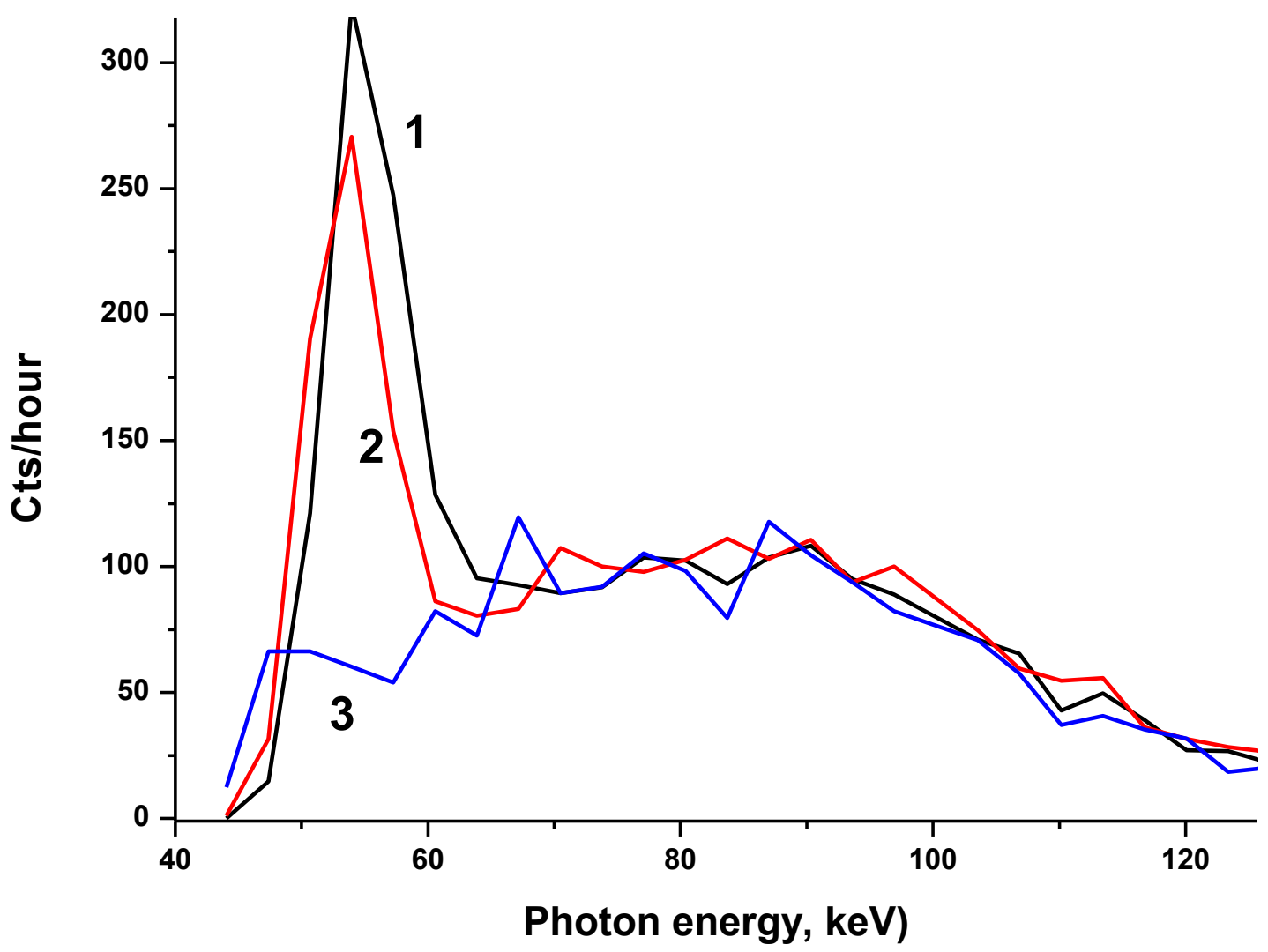

Fig. 1. Low-energy part of the spectrum of the sample of $UO_2Cl_2$ in $D_2O$ with Au NPs during laser exposure (1), immediately after the exposure (2), and 16 hours after the exposure. Data points are the discrete channels of the scintillation spectrometer. Acquisition time of each spectrum was 1 hour. Exposure to a Cu vapor laser radiation.

The signal at 54 keV increases for 1.5-2 hours during the laser exposure. Its evolution in time and after the end of the laser exposure is presented in Fig. 2.

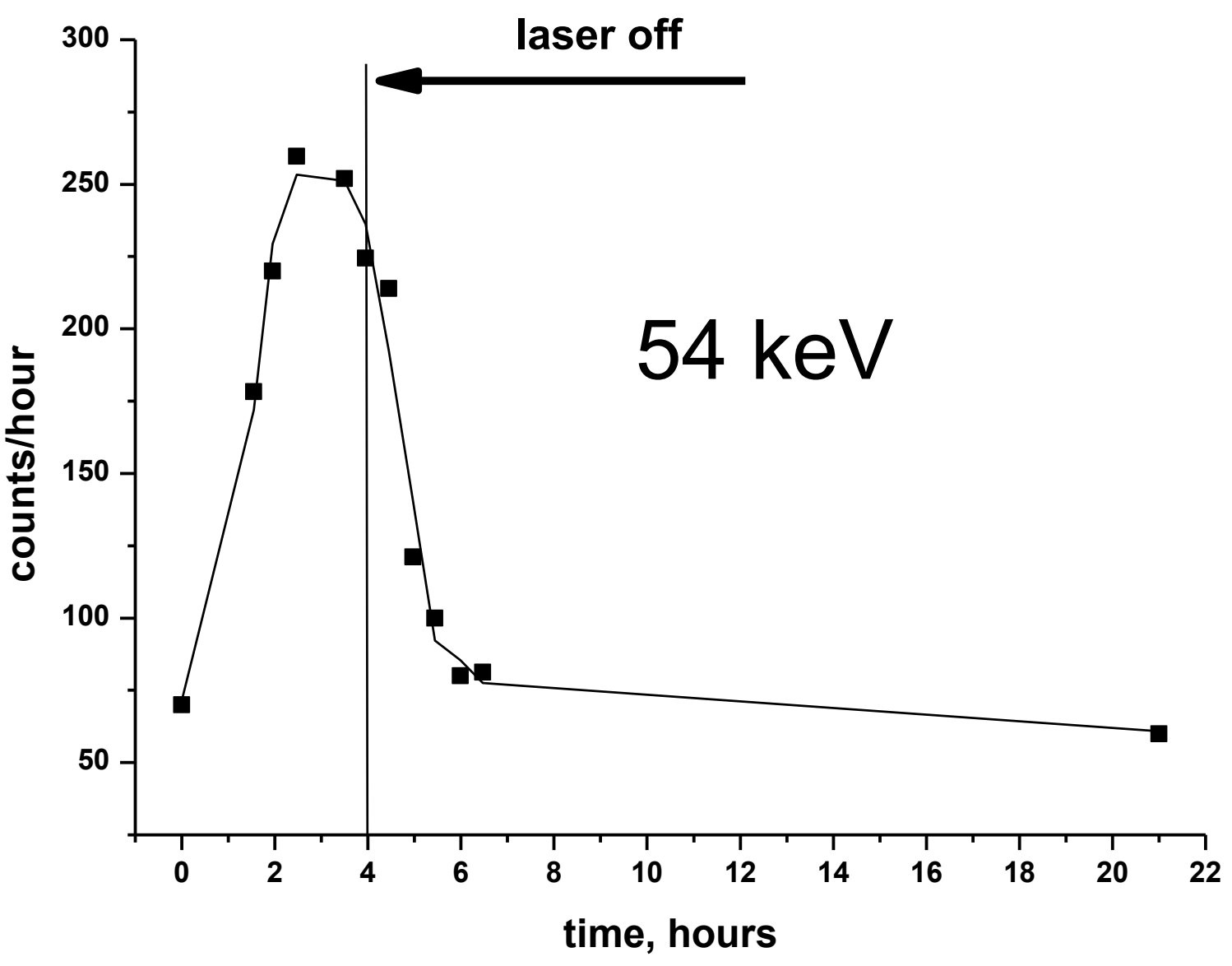

Fig. 2, a

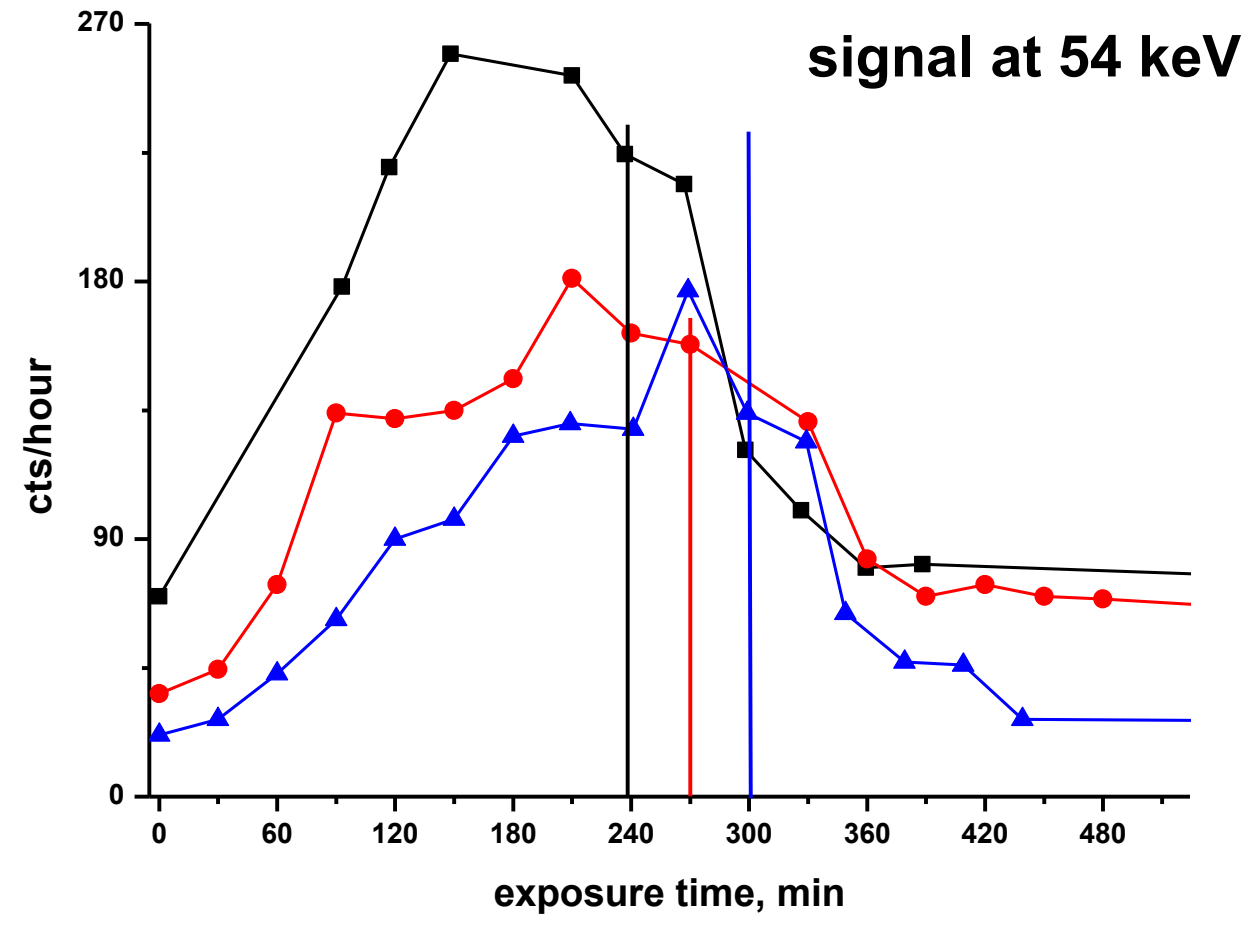

Fig. 2, b

Fig. 2. Kinetics of the gamma-emission at 54 keV photon energy. The acquisition time in each data point was 30 min (a). Different realizations of 54 keV signal counting. Vertical lines indicate the moment of shutting down the laser beam (b).

One can see that the maximum of activity at 54 keV exceeds the background level by a factor of 5. Similar emission is also observed under exposure of the same salt with Au NPs in $H_2O$. The decay of activity at this photon energy after shitting down the laser beam is presented in Fig. 3.

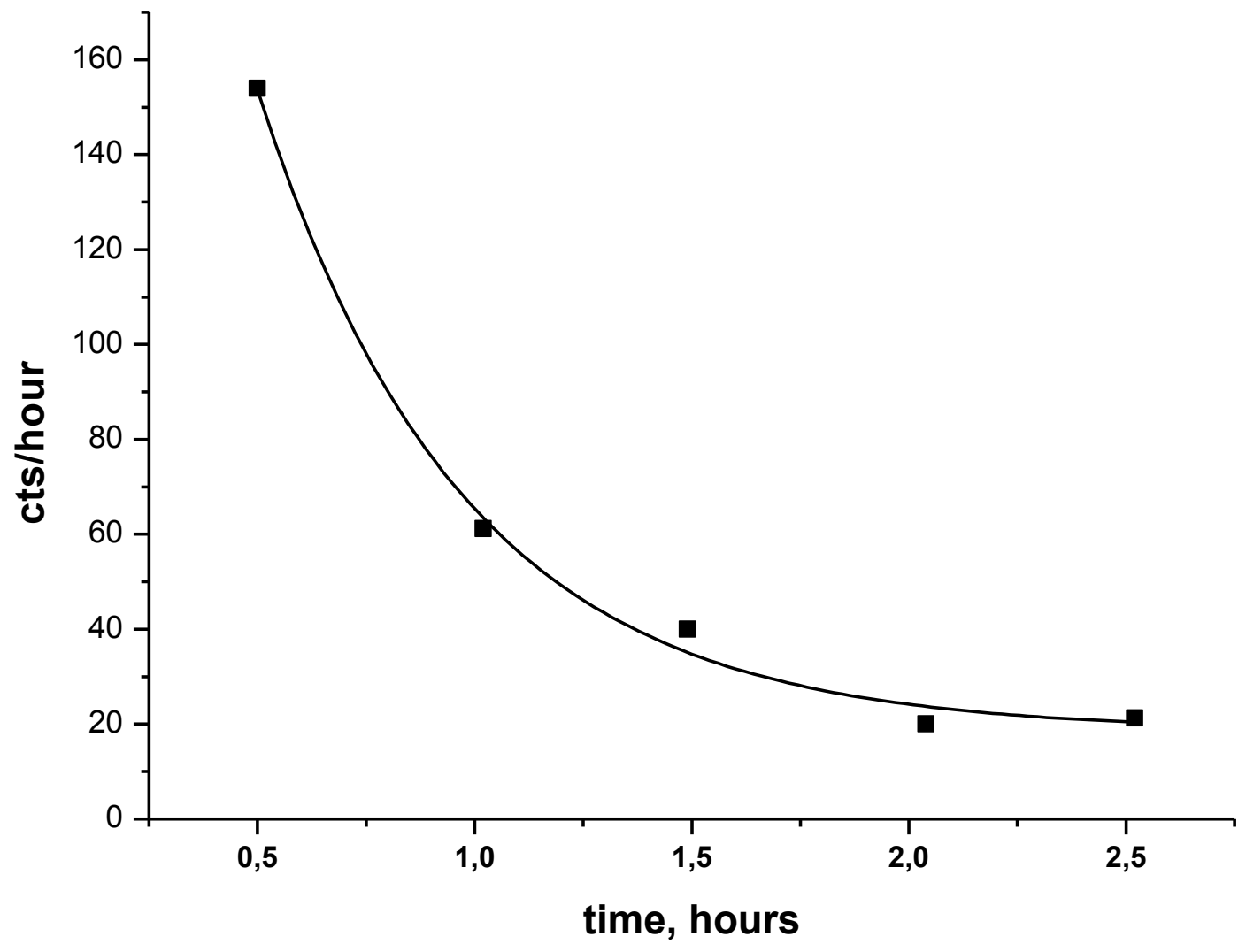

Fig. 3. Decay of activity of the sample of $UO_2Cl_2$ with Au NPs in $H_2O$ after the end of laser exposure.

This decay is well approximated with exponential dependence with half-time of 28.1 min. According to reference data, this time constant is in good agreement with the half-time of decay of the lead isotope $^{214}$Pb of 26.8 min. This isotope stands at the end of $^{238}$U branching.

Laser exposure leads to irreversible changes of the activity of nuclides in the sample. Difference spectra in the range of the detector sensitivity are presented in Fig. 4.

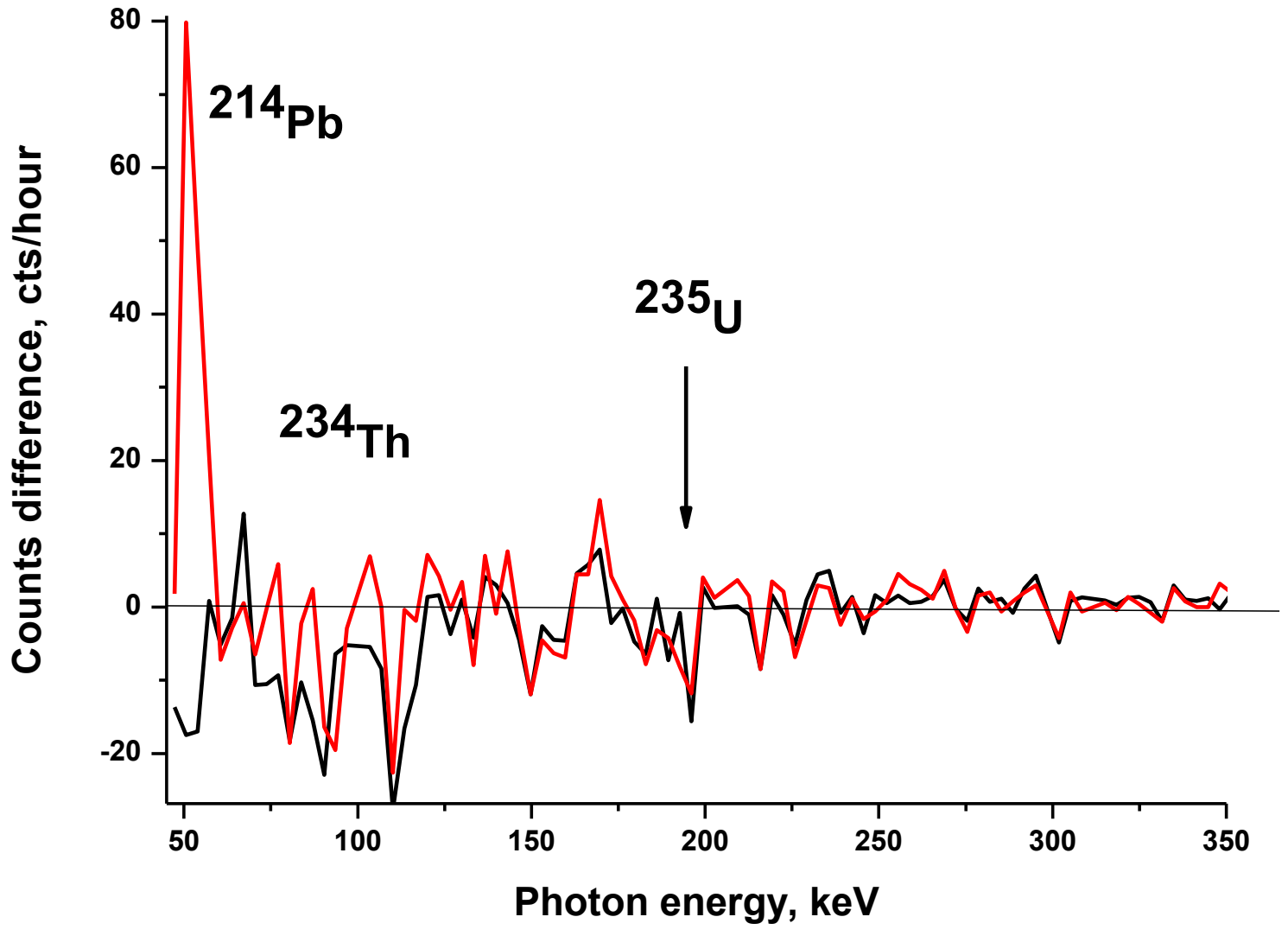

Fig. 4. Difference gamma-spectra of the sample of Au NPs in $H_2O$. Real-time spectrum minus spectrum before the laser exposure (red). Spectrum 35 hours after the laser exposure minus spectrum before the laser exposure (black).

Along with the peak of $^{214}$Pb, irreversible changes are observed near the peaks of $^{234}$Th (92.5 keV) and $^{235}$U (keV).

These data are corroborated with the gamma-analysis of samples before and after laser exposure obtained with a semiconductor gamma-spectrometer with high resolution. Fig. 5 shows these results for main nuclides of the samples.

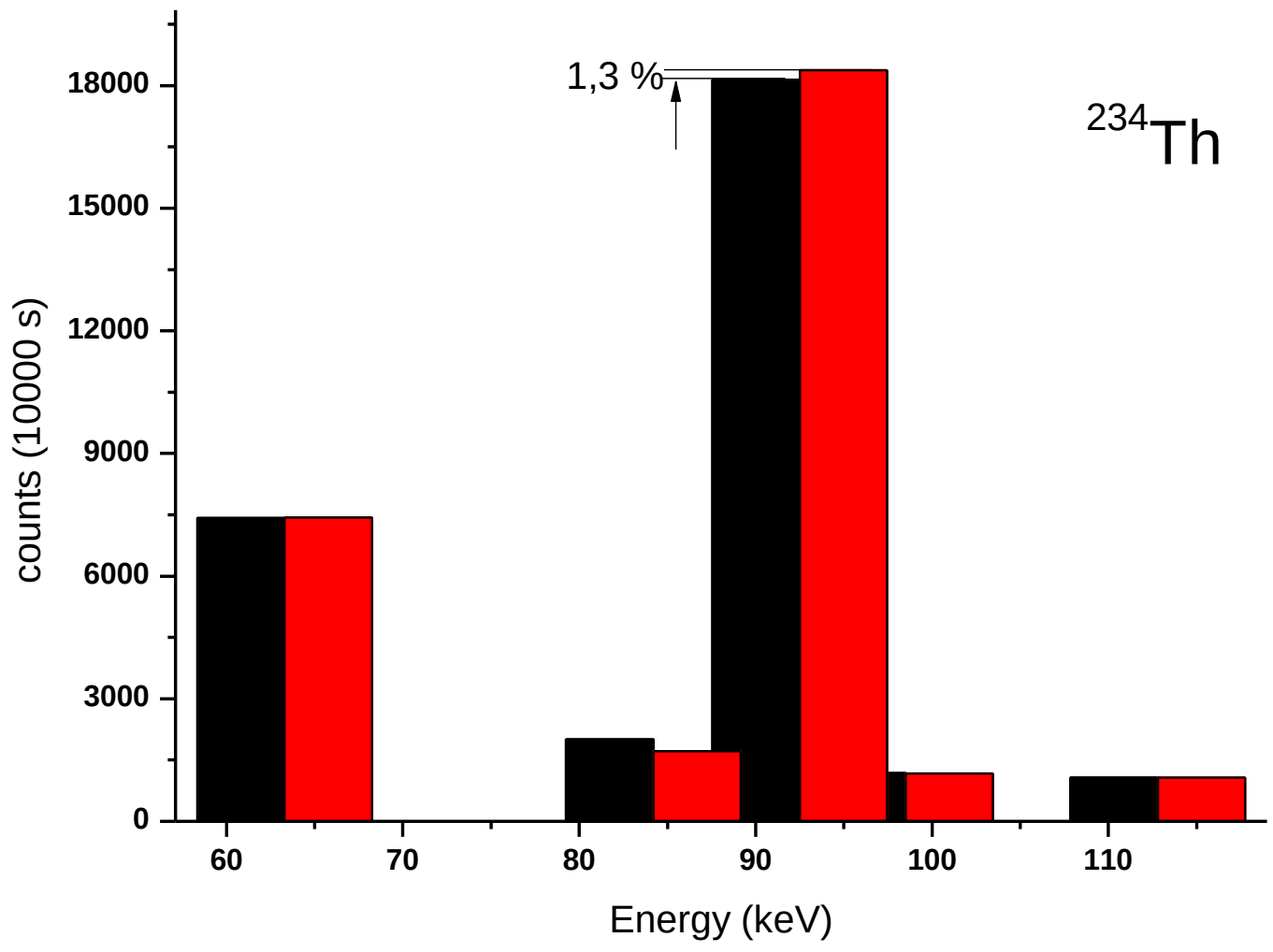

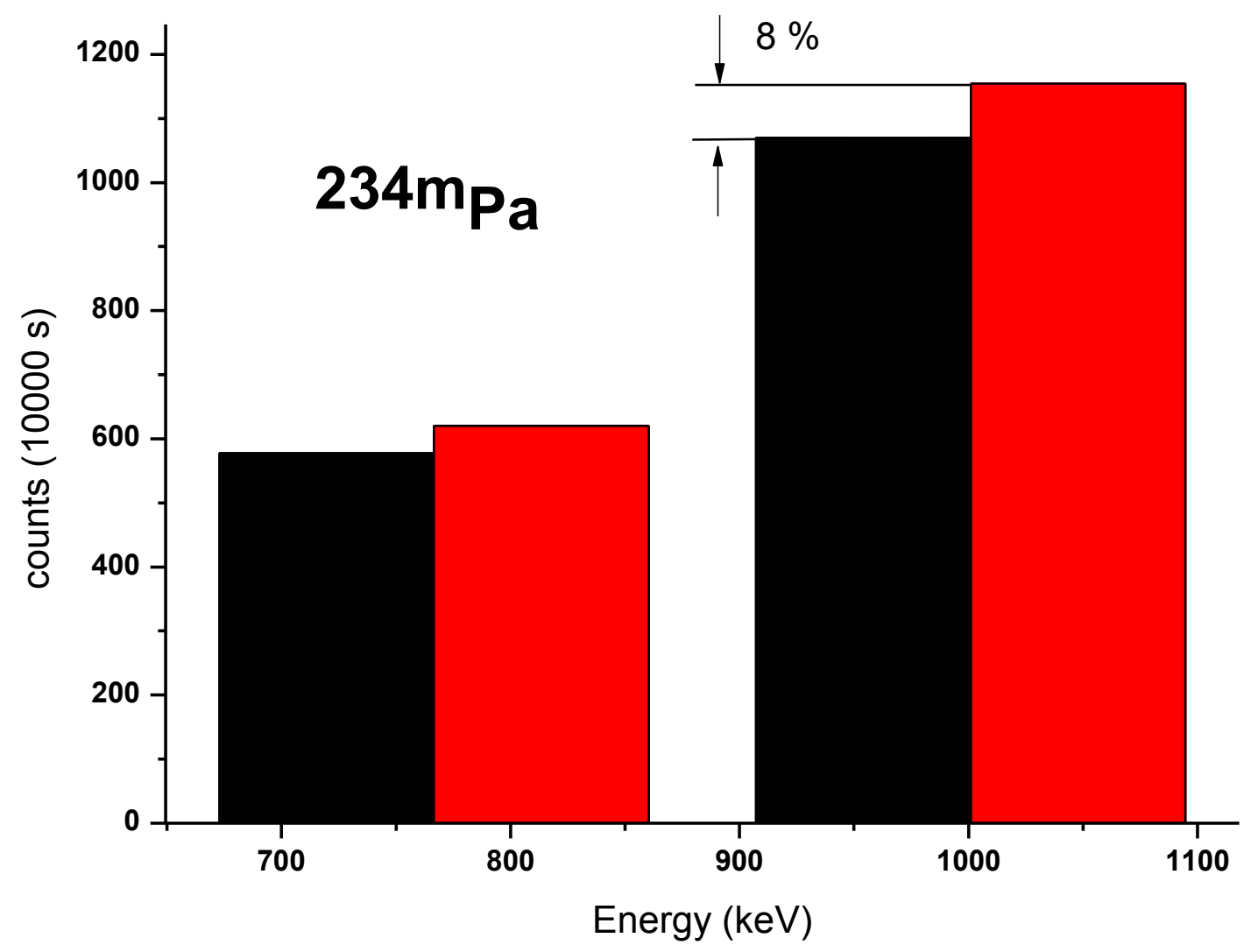

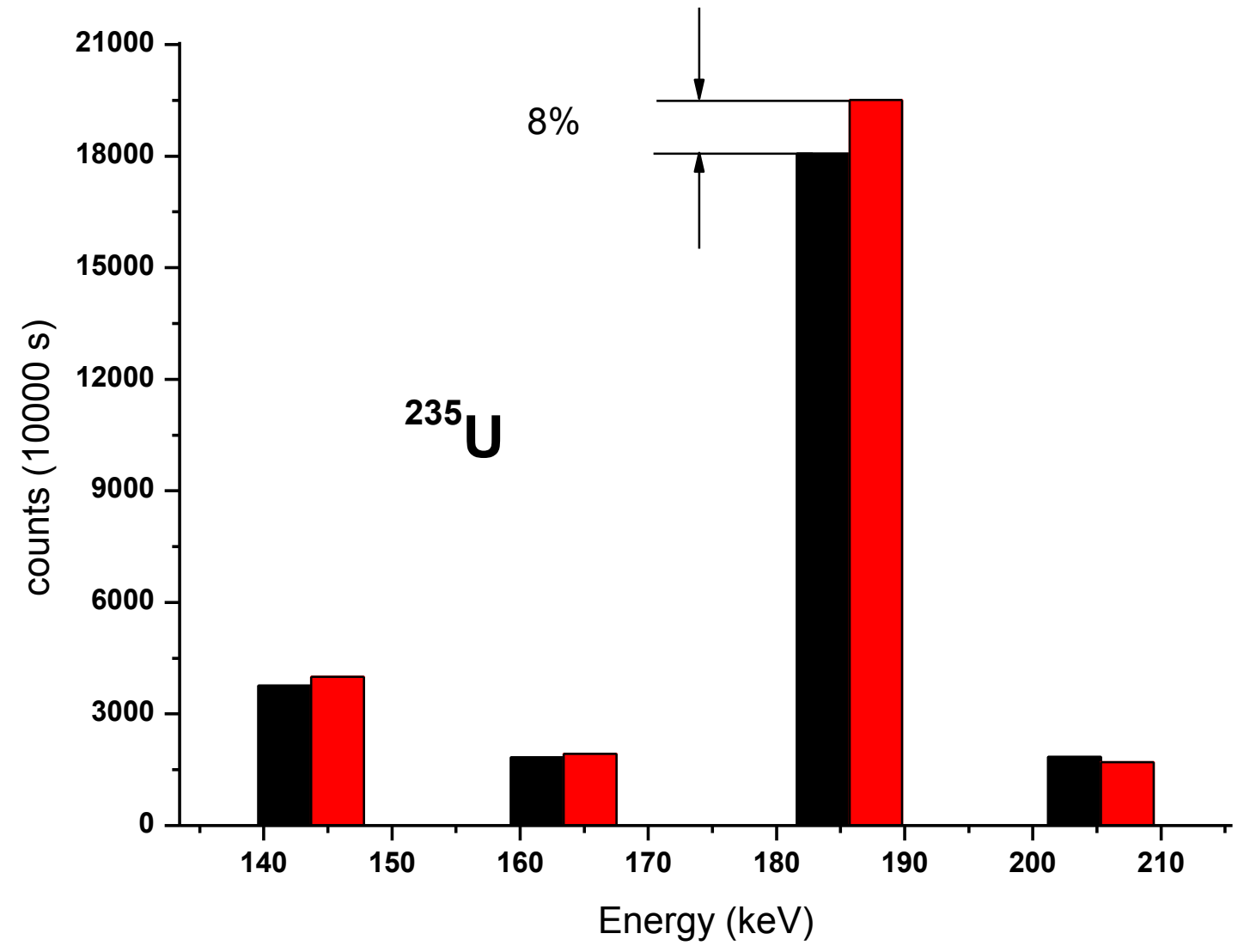

Fig. 5. Change of activity of $^{234}$Th, $^{234m}$Pa, and $^{235}$U in their main peaks after exposure of Au NPs in the solution of $UO_2Cl_2$ in $H_2O$ to radiation of a Cu vapor laser. Black columns correspond to the initial sample, red one to laser-exposed.

The difference gamma-spectrum of the samples in $D_2O$ exposed to the radiation of a Cu vapor laser is shown in Fig. 6. The difference is observed only in the vicinity of the peak of $^{214}$Pb (54 keV), while the changes of activity of both $^{234}$Th and $^{235}$U are negligible. This result is in good agreement with previous observations [3] at this level of laser intensity.

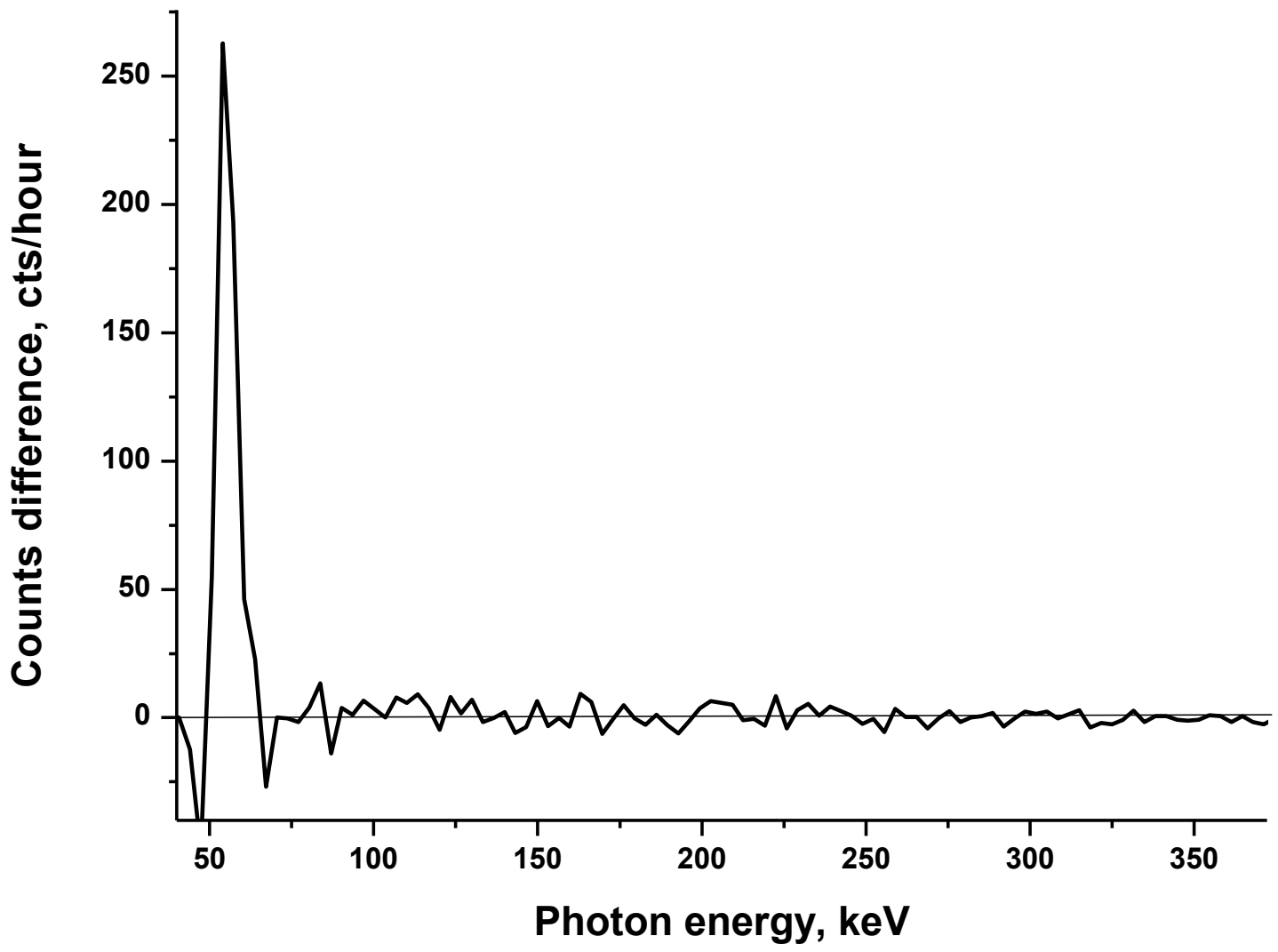

Fig. 6. Difference gamma-spectrum the sample of Au NPs in the solution of $UO_2Cl_2$ in $D_2O$ (real-time minus the spectrum 16 hours after laser exposure).

$^{214}$Pb belongs to $^{238}$U branching and its enhanced decay should be ascribed to the generation of fast neutrons during laser exposure of Au NPs in aqueous medium. This consideration stipulated the experiments on neutron counting during exposure of the samples to radiation of a Cu vapor laser. It was found that the flux of thermal neutrons (energy less than 0.4 eV) is around the background level within the accuracy of measurements. On the contrary, the flux of rapid neutrons (from 1 to 1.4×10 keV) exceeds the background level and changes during laser exposure in a non-monotonous way (see Fig. 7).

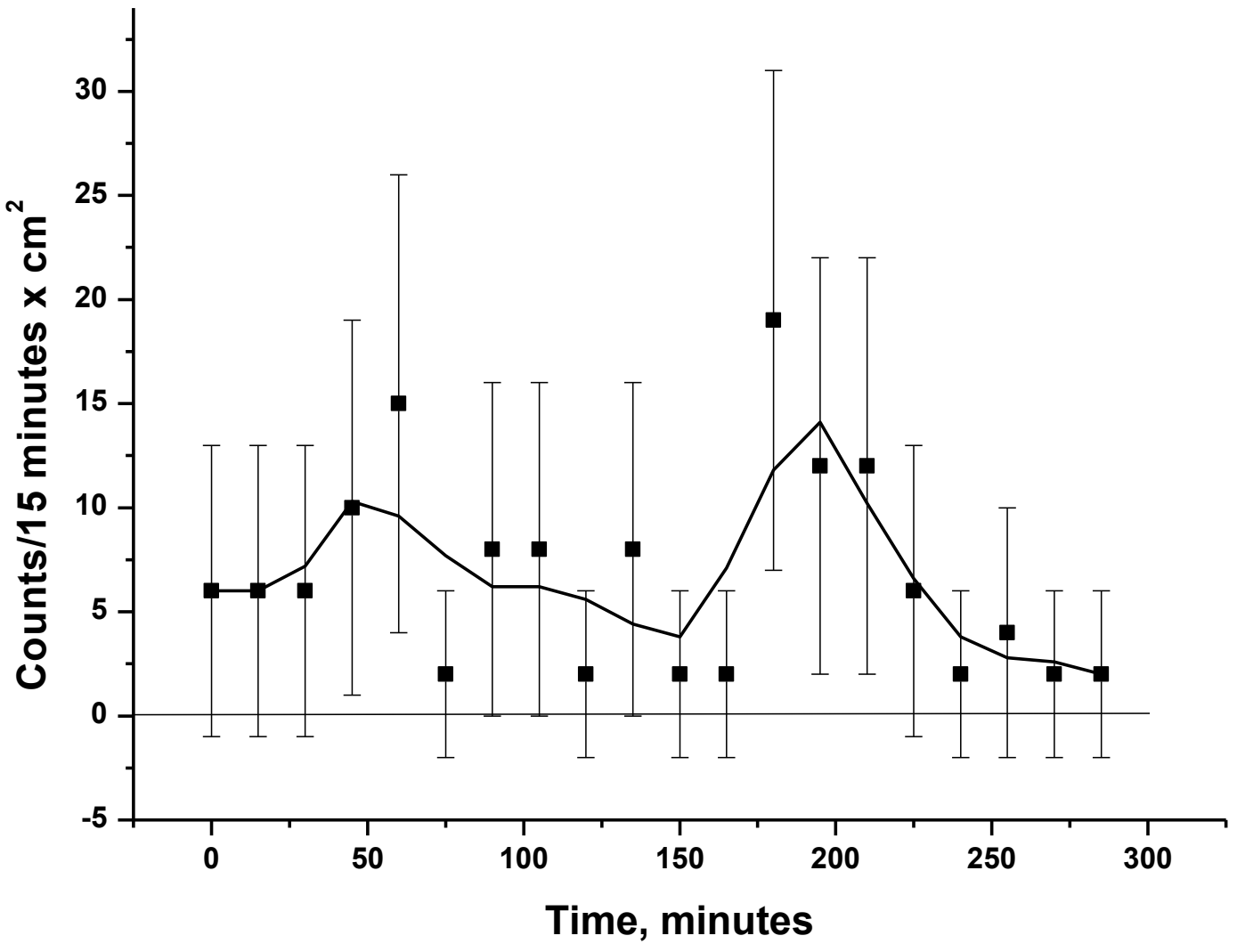

Fig. 7. Kinetics of counts of intermediate and fast neutrons measured during exposure of Au NPs in the solution of $UO_2Cl_2$ in $H_2O$ to radiation of a Cu vapor laser. Acquisition time for each data point was 15

minutes. The curve is weighted average for data points. The neutron flux was registered with a scintillation detector through $^3$He neutron moderator.

In the geometry of the experiment, the values shown in Fig. 7 correspond to the total 500 neutrons into space angle of $4\pi$ during 15 minutes of acquisition.

The activity of the samples at 54 keV reaches its maximum after 2-3 hours after the onset of the laser exposure (see Fig. 2). This is due to the modification of the absorption spectrum of the solution. Namely, the peak of the plasmon resonance of Au NPs disappears upon laser exposure, and coupling of the laser radiation to the solution becomes less efficient. Presumably, these modifications of the absorption spectrum are due to the formation of the alloyed Au-U NPs. This leads to the blue shift of the plasmon resonance of alloyed NPs compared to the initial Au NPs. For this reason, the next series of experiments was carried out using the radiation of a third harmonics of a Nd:YAG laser at wavelength of 355 nm and pulse duration of 150 ps.

The optical density of the colloidal solutions of Au NPs in both $H_2O$ and $D_2O$ is presented in Fig. 9. One can see that the starting absorption of Au NPs at 532 and 355 nm (second and third harmonics of a Nd:YAG laser) are virtually the same. However, the cross section of alloyed Au-U NPs at 355 nm is much higher, and therefore, coupling of the laser radiation will improve upon laser exposure of the samples. In other words, the system “NPs – laser radiation” is characterized by a positive feedback at this wavelength.

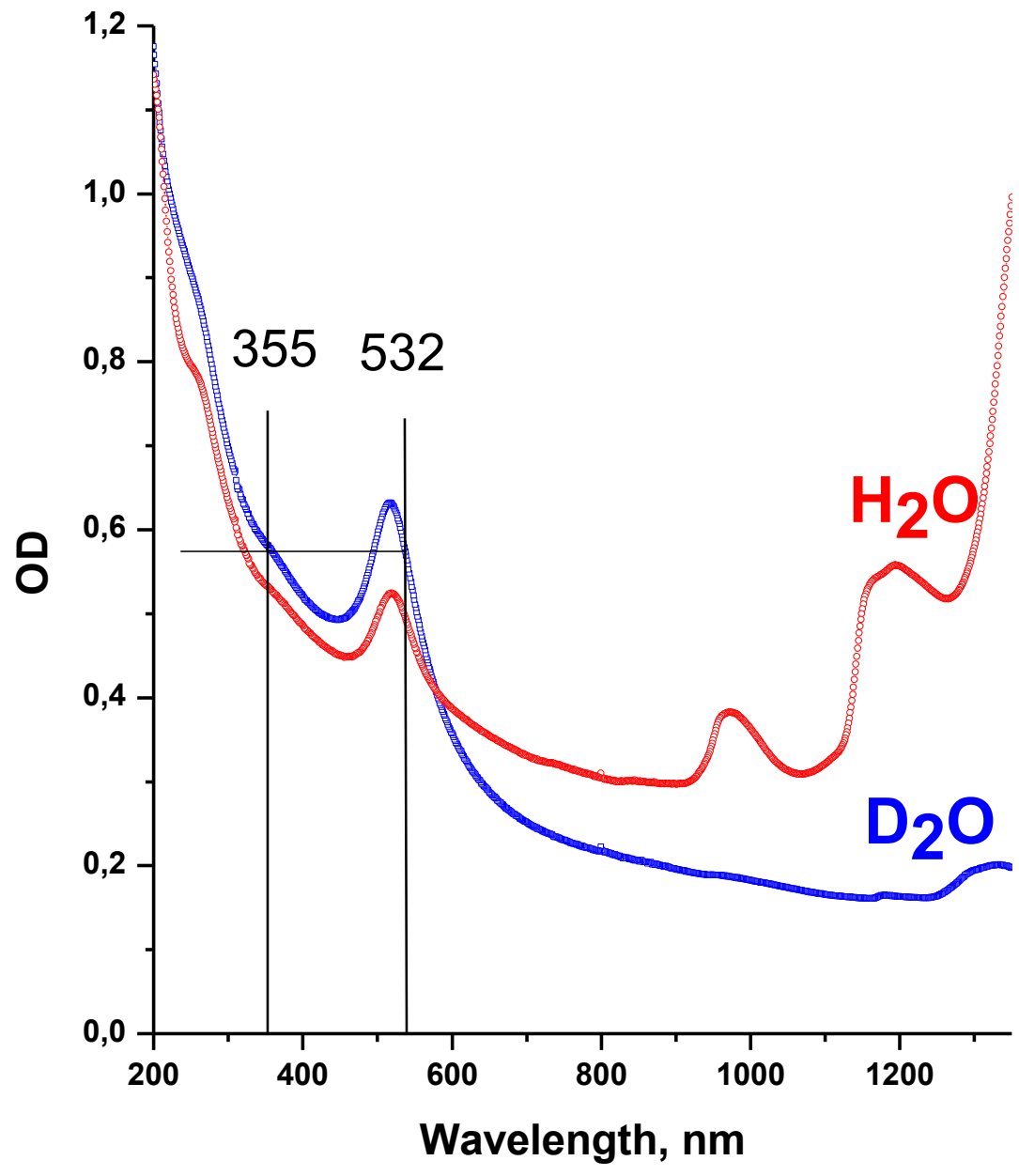

Fig. 8. Optical density of Au NPs in $H_2O$ and $D_2O$ from near IR to UV region. Absorption peaks in the near IR region are due to obertones of valence oscillations of both waters. Au NPs were obtained by ablation of a gold target by radiation of a Cu vapor laser.

In Fig. 9 the activity of main nuclides in the samples are plotted as the function of the concentration of $UO_2Cl_2$ in the colloidal solution of Au NPs. The total volume of the liquid was kept at 2 ml, and only the ratio of concentrations of Au NPs and $UO_2Cl_2$ was varied.

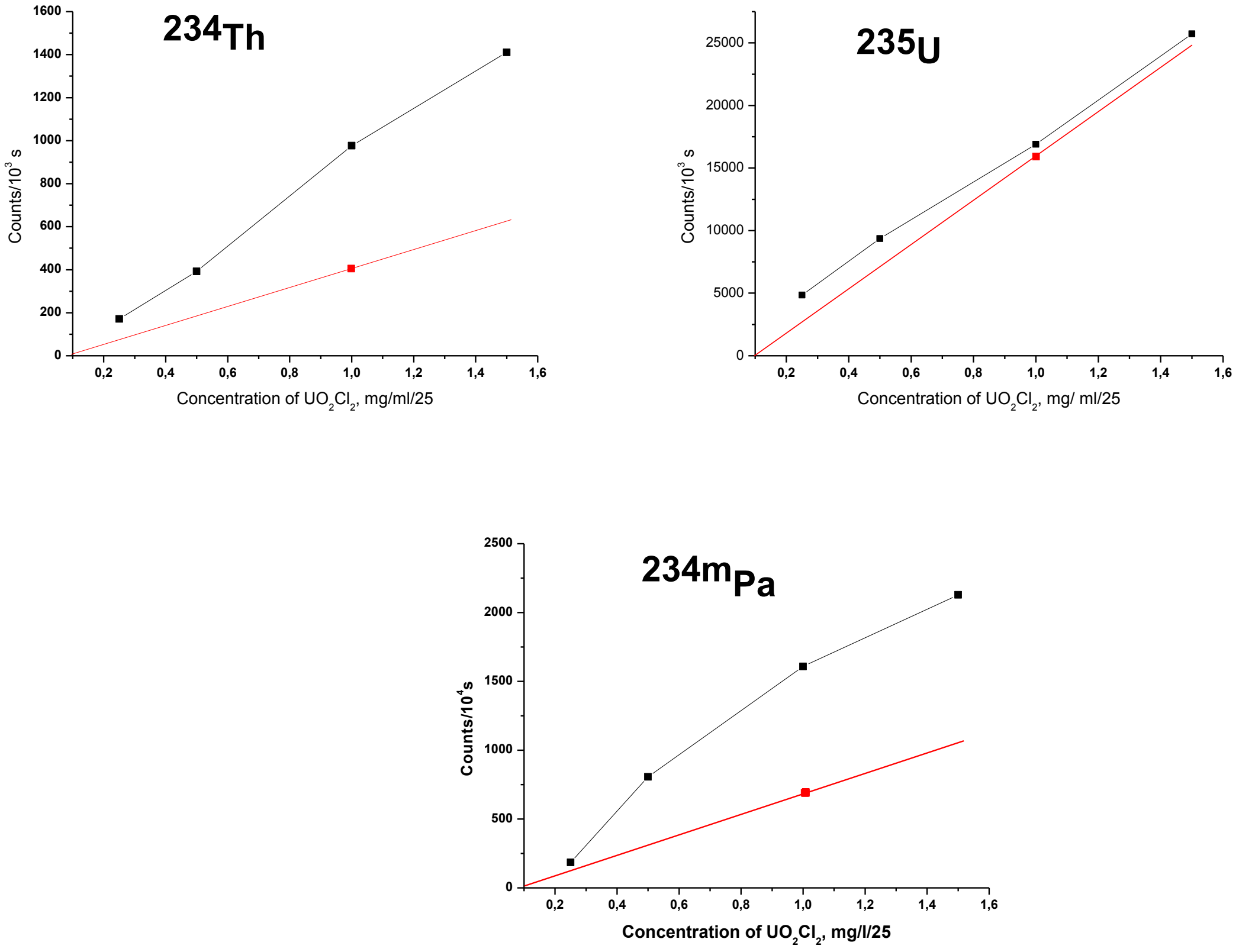

Exposure in $H_2O$

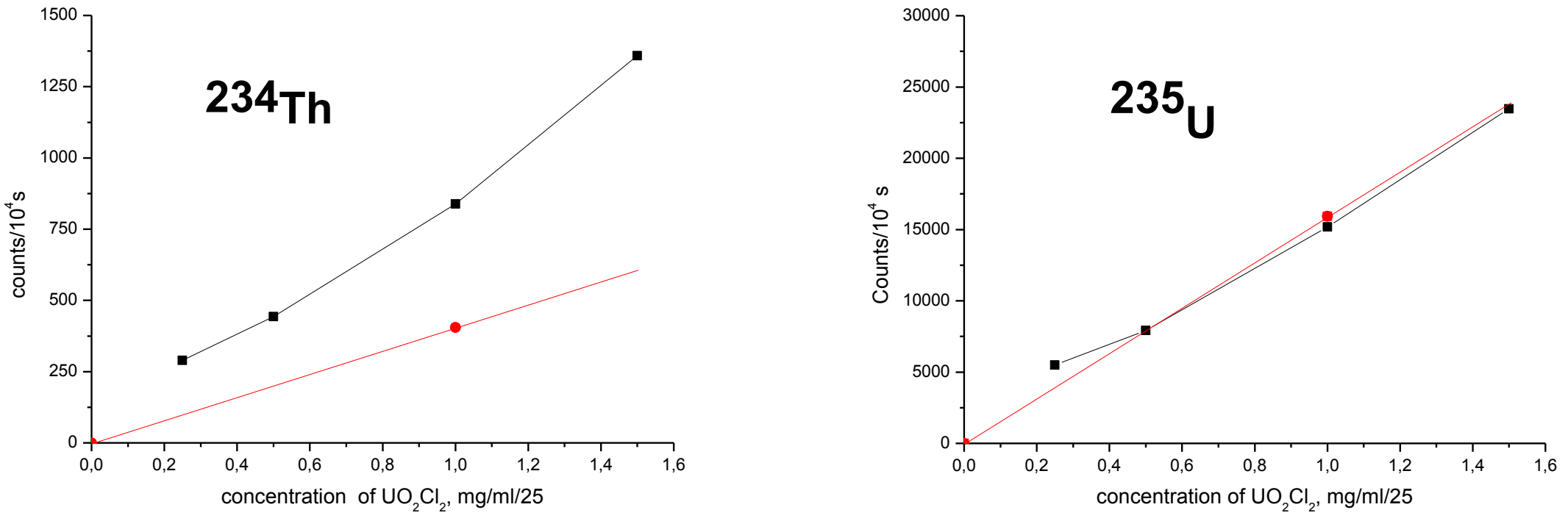

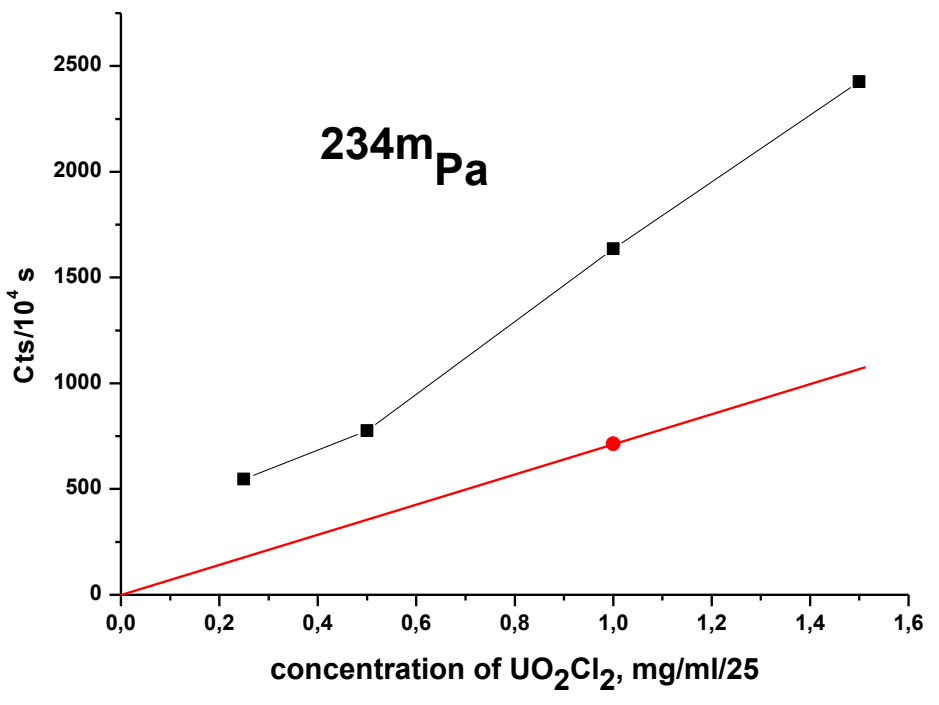

Exposure in $D_2O$

Fig. 9. Dependence of the activity of $^{234}$Th and $^{235}$U on the concentration of $UO_2Cl_2$ in the colloidal solution of Au NPs. Straight red line is the dependence of respective nuclides in the non-exposed solution. Third harmonics of a Nd:YAG laser at 355 nm, estimated peak intensity in the liquid of $10^{13}$ W/cm$^2$. Each sample was subjected to 36 000 laser pulses.

One can conclude that at this level of laser intensity the difference between waters, $H_2O$ and $D_2O$ is not very pronounced. The increase of $^{234}$Th content is observed at any concentration of U salt, and this increase is supra-linear with concentration. Note that the half-life time of $^{234}$Th is 24 days, while measurements of activity of samples were carried out 30 days after laser exposure. This means that the actual increase of $^{234}$Th content is even higher and can be estimated by a factor of 4. Similarly, the increase of content of short-living nuclide $^{234}Pa_m$ is observed above its equilibrium value. These two nuclides stand in the branching series of $^{238}$U: $^{238}U \rightarrow {}^{234}Th \rightarrow {}^{234}Pa_m \rightarrow$ etc. Therefore, the increase of $^{234}$Th content is due to the accelerated decay of $^{238}$U. And this decay is highly selective towards $^{238}$U isotope, since the content of $^{235}$U remains within that of the initial solution, as one can see in Fig. 10. $^{214}$Pb could not be detected for these samples due to its short decay time.

It is pertinent to note that the ion $UO_2^-$ is decomposed under the exposure to UV laser radiation. This can be deduced from the disappearance of its green luminescence well visible in the early stages of exposure.

In the conditions of secular equilibrium, the ratio of the contents $N_i$ of neighboring nuclides is equal to the ratio of their decay constants $T_{(1/2)i}$:

$$N_1/N_2 = T_{(1/2)1}/T_{(1/2)2}$$

The half-life time of $^{238}$U and $^{234}$Th is $10^9$ years и 24 days, respectively. Laser exposure of the samples increases the thorium content for at least one order of magnitude taking into account that thorium itself undergoes accelerated decay into $^{234m}$Pa under laser exposure. The estimations show that around $10^6$ nuclei of $^{238}$U decay during one laser pulse at 355 nm wavelength and pulse duration of 150 ps. It should be noted that the increase of the activity of $^{234}$Th is super-linear, which might be the indication of the fact that neutrons released during the decay of $^{238}$U may initiate the decay of other nuclei.

The activity of another isotope, $^{235}$U behaves itself differently under laser exposure of the solutions. As one can see in Fig. 10, its activity does not change after laser exposure except for low U content, and $^{235}$U content remains constant. Since $^{238}$U decays, one may consider this behavior as an enrichment of U with $^{235}$U, though in this case these isotopes are not separated. It is $^{238}$U which is decayed selectively. This selectivity may be explained by mean energy of neutrons that are generated under laser exposure of Au NPs at different peak laser power in the medium. The content of $^{235}$U increased under exposure under laser exposure in $H_2O$ and decreased under exposure in $D_2O$ at laser peak intensity of $10^{10}$W/cm$^2$ at wavelength of 510 nm and pulse duration of 10 ns [3].

Deviation of the activity of $^{234}$Th from its equilibrium value has been reported under electric explosion of a Ti foil in the solution of uranil-nitrate in $H_2O$ [4]. The authors observed 30% decrease of the concentration of this nuclide immediately after the explosion and its slow restoration to the equilibrium value after 24 days (?) after the experiment. Qualitatively similar results are observed in the conditions of the present work. Fig. 10 presents results on the activity of $^{234}$Th before, during, and after the laser exposure of the sample of Au NPs in $H_2O$ to radiation of a Cu vapor laser.

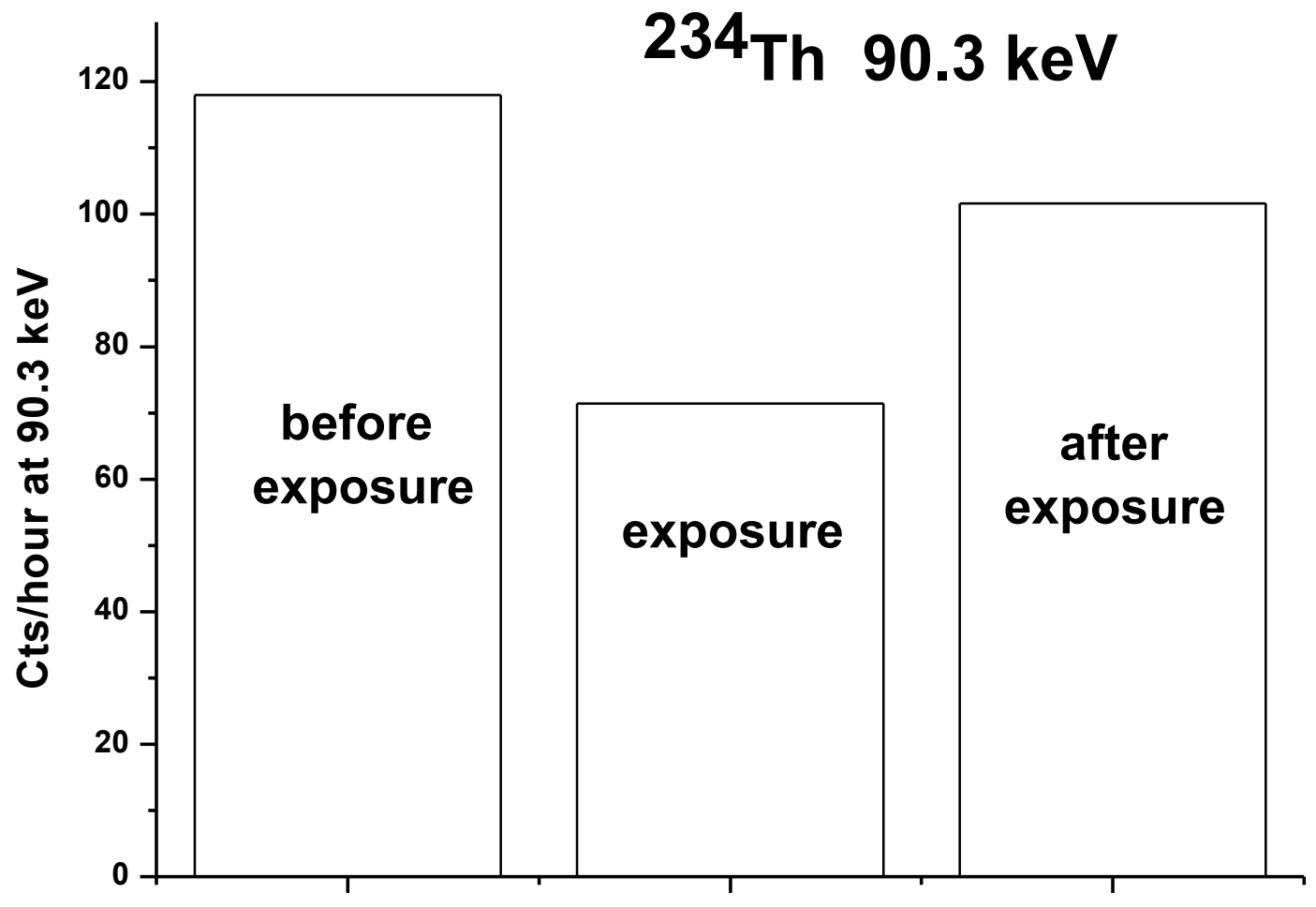

Fig. 10. Normalized activity of the sample of Au NPs in $H_2O$ with $UO_2Cl_2$ at the photon energy of 90.3 keV before, during, and 2.5 hours after exposure of the sample to radiation of a Cu vapor laser.

One can see that the activity of this nuclide indeed decreases upon laser exposure but recovers to almost its initial value in several hours. In any case such a behavior is intensity-dependent, since at higher intensity the activity of $^{234}$Th increases after laser exposure (see Fig. 10). The increase of the activity of this nuclide after the end of the laser exposure is possible only if the accelerated decay of $^{238}$U continues without laser irradiation. Natural decay of $^{238}$U would not provide the restoration of the $^{234}$Th content to its equilibrium value in several hours. The presence of nuclear reactions that continue in the colloidal solution after the end of the laser exposure was reported in our previous work with $^{232}$Th salt [2]. The saturation of the liquid and therefore metallic NPs with either $H_2$ or $D_2$ increases the rate of these transformations, as it was reported for laser exposure of Au NPs with $^{238}$U salt in water [3].

The two approaches, either laser exposure of metallic NPs in liquids or electro-explosion of a foil in liquid are not so much different as one can believe it at the first sight. Indeed, electro-explosion of foils or wires is well-known method of generation of nanoparticles. These NPs should contain hydrogen due to partial dissociation of water during the dispersion of the foil in water. Thus, nanometric size of metallic species and their saturation with hydrogen are the common point between these two processes. However, this does not explain the underlying mechanisms that can trigger accelerated decay of radio-nuclides. This point requires further studies.

In conclusion, it is found that laser exposure of gold nanoparticles in aqueous solutions of $UO_2Cl_2$ leads to accelerated decay of $^{238}$U nuclei and deviation from secular equilibrium of its daughter elements. The samples demonstrate the enhanced gamma-emission in the range of 54 keV under laser intensity of order of $10^{11}$ W/cm$^2$ and pulse duration of 10 ns. This gamma-emission is identified as the synthesis and decay of short-living isotope $^{214}$Pb. Deviation from equilibrium in concentration of nuclides of $^{238}$U branch is especially pronounced at laser peak power of $10^{12}$ W/cm$^2$ at the wavelength of 355 nm and pulse duration of 150 ps. The content of both $^{234}$Th and $^{234m}$Pa increases by a factor of 4 compared to equilibrium value as a result of laser exposure of the colloidal solutions. The content of $^{235}$U in these conditions remains virtually the same.

Acknowledgements

The work was partially supported by Russian Foundation for Basic Research, grants ## 07-02-00757, 08-07-91950, and by Scientific School 8108.2006.2. Dr. A.V. Goulynin is thanked for gamma-measurements and helpful discussions.